\documentclass[conference]{IEEEtran}
\IEEEoverridecommandlockouts
\usepackage{cite}
\usepackage{amsmath,amssymb,amsfonts}
\usepackage{algorithmic}
\usepackage{graphicx}
\usepackage{svg}
\usepackage{multirow}
\usepackage{array}

\usepackage{makecell} % Allows multi-line column headers
\usepackage{stfloats} % Add this to your preamble
\usepackage{ragged2e}
\usepackage{lipsum} % For dummy text
\usepackage{comment}
\usepackage{afterpage}
\usepackage{float}
\usepackage{textcomp}
\usepackage{xcolor}
\usepackage{hyperref}
\usepackage{url}
\usepackage{enumitem}
\usepackage{multirow}
\usepackage{array}
\usepackage{adjustbox}

\renewcommand{\arraystretch}{1.5} % Line spacing between rows

\def\BibTeX{{\rm B\kern-.05em{\sc i\kern-.025em b}\kern-.08em
    T\kern-.1667em\lower.7ex\hbox{E}\kern-.125emX}}
\begin{document}

% \title{NVDLA and RISC-V Based Bare-Metal SoC Design for Edge AI
% \thanks{This work was partly funded by Taighde Éireann – Research Ireland  through the Research Ireland Centre for Research Training in Machine Learning (18/CRT/6183)}
% }

%  \thanks{Corresponding Author: Vineet Kumar}

% \author{
%     \IEEEauthorblockN{ Vineet Kumar, Ajay Kumar M, Yike Li, Deepu John, Shreejith Shanker}
%     \IEEEauthorblockA{
%         \textit{School of Electrical and Electronic Engineering} \\
%         \textit{University College Dublin} \\
%         Dublin, Ireland \\
%         vineet.bitsp@gmail.com, ajay.kumarm@ucdconnect.ie, yike.li@ucdconnect.ie, deepu.john@ucd.ie, shankers@tcd.ie 
%     } 
% }

% \maketitle
\title{Bare-Metal RISC-V + NVDLA SoC for Efficient Deep Learning Inference}

\author{
    Vineet Kumar\IEEEauthorrefmark{0}\thanks{Corresponding Author: Vineet Kumar. This work was partly funded by Taighde Éireann – Research Ireland through the Research Ireland Centre for Research Training in Machine Learning (18/CRT/6183)},
    Ajay Kumar M\IEEEauthorrefmark{0},
    Yike Li\IEEEauthorrefmark{0},    
    Shreejith Shanker\IEEEauthorrefmark{2},
    Deepu John\IEEEauthorrefmark{0}\\[+0.6em]
    \IEEEauthorblockA{\IEEEauthorrefmark{0}School of Electrical and Electronic Engineering, University College Dublin, Dublin, Ireland}\\[-0.6em]
    \IEEEauthorblockA{\IEEEauthorrefmark{2}Department of Electronic and Electrical Engineering, Trinity College Dublin
    , Dublin, Ireland}\\[-0.6em]
    \IEEEauthorblockA{vineet.bitsp@gmail.com, ajay.kumarm@ucdconnect.ie, yike.li@ucdconnect.ie, shankers@tcd.ie, deepu.john@ucd.ie}
}

\maketitle

\begin{abstract}

This paper presents a novel System-on-Chip (SoC) architecture for accelerating complex deep learning models for edge computing applications through a combination of hardware and software optimisations. The hardware architecture tightly couples the open-source NVIDIA Deep Learning Accelerator (NVDLA) to a 32-bit, 4-stage pipelined RISC-V core from Codasip\textsuperscript{\textregistered} called $\mu$RISC\_V. To offload the model acceleration in software, our toolflow generates bare-metal application code (in assembly), overcoming complex OS overheads of previous works that have explored similar architectures. This tightly coupled architecture and bare-metal flow leads to improvements in execution speed and storage efficiency, making it suitable for edge computing solutions. We evaluate the architecture on AMD’s ZCU102 FPGA board using NVDLA-small configuration and test the flow using LeNet-5, ResNet-18 and ResNet-50 models. Our results show that these models can perform inference in 4.8 ms, 16.2 ms and 1.1 s respectively, at a system clock frequency of 100 MHz.
\end{abstract}


\section{Introduction}

The growing computational demands of AI workloads and the limitations of edge devices have driven the need for specialized hardware accelerators. The rise of open-source hardware has enabled the development of accelerators like the NVIDIA Deep Learning Accelerator (NVDLA)~\cite{b1,OSH,RV,b2}. NVDLA is a scalable, configurable, open-source inference engine suited for edge AI. Its integration with RISC-V presents a compelling solution for deep learning acceleration, as explored in several studies.

Previous works \cite{b3,b4,b5} have examined the integration of RISC-V with NVDLA to enhance deep learning inference efficiency and flexibility. However, these studies primarily focus on simulation-based implementations rather than real hardware deployments, such as FPGAs. In \cite{b6},  an FPGA-based prototype incorporating multiple instances of NVDLA and a RISC-V core is presented, but details on resource utilization and integration methodologies are not provided. Additionally, these studies \cite{b3,b4,b5,b6} rely on a Linux-based kernel to execute neural network models, requiring NVDLA drivers and resulting in significant software overhead.
Few works \cite{b8, b9, b10} have demonstrated FPGA-based implementations of NVDLA integrated with existing processor cores in SoCs, often utilizing Linux-based environments such as PetaLinux \cite{b8}. Other works \cite{b11,b12,b13,b14} have explored the use of NVDLA in various research applications but without a focus on integrating the accelerator with a RISC-V core. The reliance on Linux kernel for executing deep learning workloads introduces additional performance and storage overhead, making these solutions less suitable for resource-constrained edge devices.

In this paper, we present the design of an open-source NVDLA and RISC-V based SoC, which takes a neural network model as input and executes it on NVDLA using RISC-V assembly code without relying on a Linux kernel. Moreover, the SoC is demonstrated on FPGA by running neural network models. Instead of using a Linux kernel-managed driver stack, we leverage configuration files (traces) to directly configure NVDLA’s registers, serving as an execution control sequence. The official NVDLA release provides pre-generated configuration files for basic hardware tests (e.g., sanity checks, convolution, and pooling layer tests). However, no guidelines are available on how these files were generated or how to create them for arbitrary neural networks. This work addresses this gap by proposing a methodology to generate configuration files for arbitrary Caffe-based neural networks. These files are then converted into RISC-V assembly code, enabling direct hardware configuration of NVDLA.
The key contributions of this work include:
\begin{itemize}
    \item[--] \textit{Design of an SoC architecture based on NVDLA and RISC-V and its implementation on FPGA}
    \item[--] \textit{Automated generation of configuration files and weight extraction for arbitrary Caffe neural network models \footnote{\url{https://github.com/vineetbitsp/riscv-nvdla-sw}} }
    \item[--] \textit{Tightly coupled hardware architecture and bare-metal assembly-based execution, eliminating the need for a Linux kernel and additional storage}
\end{itemize}

For system design, we integrate NVDLA with a Codasip~$\mu$RISC\_V core and implement the design on an AMD ZCU102 FPGA board. The system is validated using LeNet-5, ResNet-18, and ResNet-50 neural network models.

 \section{Related Works}
\label{related_work}
Several prior studies have explored the integration and evaluation of NVDLA within different computing environments, primarily focusing on simulation-based approaches and Linux-kernel based FPGA implementations.
Gem5-NVDLA \cite{b4}  serves as a valuable tool for analyzing design trade-offs and evaluating NVDLA’s performance in a simulated environment.  However, this work is limited in scope, as it does not support the  small configuration of NVDLA (nv\_small) \cite{b2}. 
Gonzalez and Hong \cite{b5} conducted a comparative study of the NVDLA and Gemmini accelerators within the Chipyard framework, assessing their respective advantages for deep learning workloads. While insightful, this work is framework-specific, restricting its applicability to Chipyard users. 

Farshchi et al.~\cite{b3} investigated the integration of NVDLA with RISC-V-based SoCs using FireSim, a cycle-accurate simulation platform, to evaluate performance in object detection tasks. However, their study is limited to simulation-based analysis and does not address the practical challenges of FPGA-based deployment or a custom ASIC design. Notably, their simulation assumes an unrealistic NVDLA operating frequency of 3.2 GHz---equivalent to the CPU clock—due to FireSim’s constraints, whereas in practical FPGA implementations, NVDLA operates at frequencies below 100 MHz~\cite{b6}. 

Giri et al. \cite{b6} proposed an open-source embedded system platform for agile heterogeneous SoC design and demonstrated FPGA-based prototypes incorporating multiple instances of NVDLA alongside the Ariane RISC-V 64-bit processor core. However, their work does not provide details on FPGA resource utilization or integration methodologies.

To the best of our knowledge, all prior works \cite{b3,b4,b5,b6,b7,b8,b9,b10} require a Linux kernel to configure and operate NVDLA. In contrast, our work employs bare-metal assembly programming to directly configure NVDLA registers for a given neural network. Furthermore, our implementation supports both small and full configurations (nv\_small and nv\_full) of NVDLA.

\begin{figure}[t!]
\centering
\includegraphics[width=0.55\linewidth]{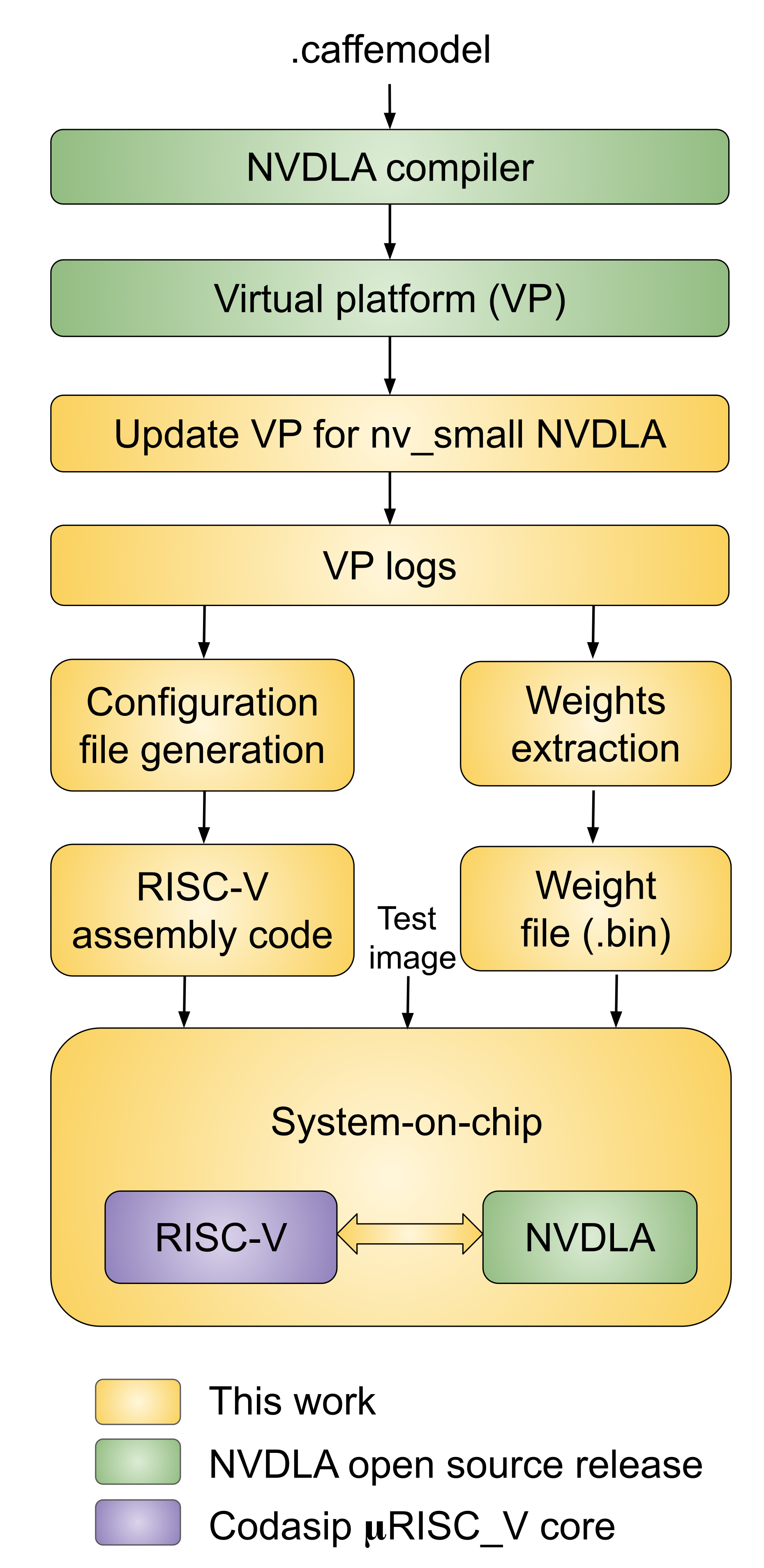}
\caption{The proposed system and software development flow.}
\label{fig:system}
\vspace{-4mm}
\end{figure}

\section{The Proposed system}

This section outlines the architecture of the proposed SoC, followed by detailed software and hardware development in the subsequent section. Fig.~\ref{fig:system} presents the software generation workflow, which converts a trained neural network model into RISC-V assembly code and a corresponding weight file. As this process is model-specific and performed only once, it is executed offline using NVDLA's virtual platform (VP) in conjunction with the software development methodology described in Section \ref{subsec:sw}. 

\begin{figure}[t!]
\centering
\includegraphics[width=0.9\linewidth]{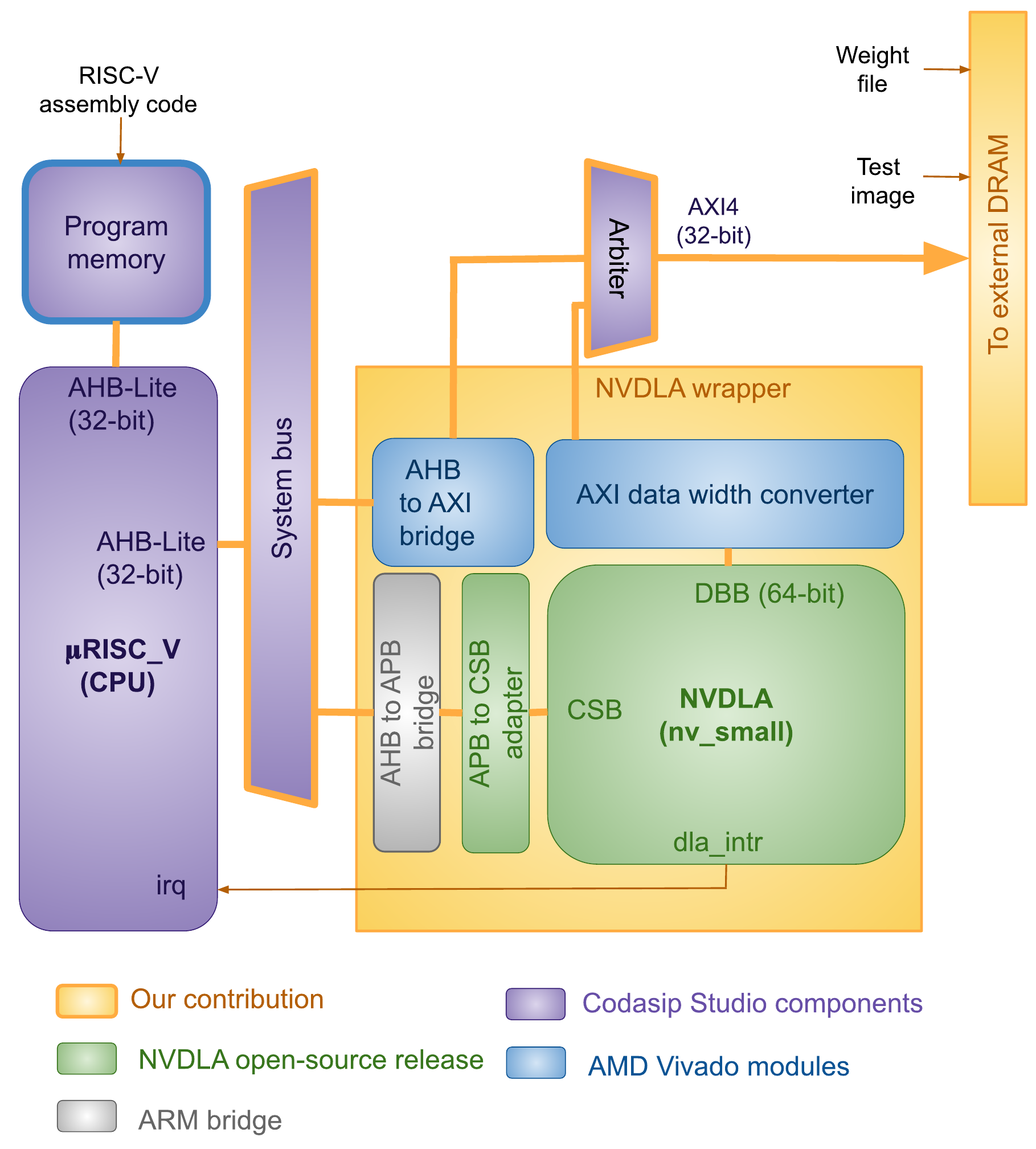}
\caption{The system-on-chip.}
\label{fig:soc}
\vspace{-6mm}
\end{figure}

Fig.~\ref{fig:soc} illustrates the architectural design of the proposed SoC. The system integrates the NVDLA accelerator with a $\mu$RISC-V core through a system bus, an arbiter, and a custom NVDLA wrapper. The system bus—comprising an internal decoder and arbitration logic—enables communication between the $\mu$RISC-V core and two memory-mapped slave devices: the NVDLA engine and DRAM-based data memory. Given the shared access to data memory, an arbiter manages potential conflicts between the core and NVDLA. 
The NVDLA wrapper encapsulates the accelerator hardware alongside interface bridges and a data width converter to address mismatches between the $\mu$RISC-V and NVDLA interfaces. Specifically, an AXI data width converter connects the NVDLA’s 64-bit data backbone (DBB) interface to the 32-bit data memory. The $\mu$RISC-V core employs an AHB-Lite interface for access to both program and data memory. Communication with NVDLA’s configuration space bus (CSB) requires an AHB-Lite to APB bridge, leveraging the existing APB-to-CSB adapter provided by the NVDLA package. The AHB-APB bridge, available as an open-source ARM design, facilitates this integration. Furthermore, an AHB-Lite to AXI bridge enables connectivity between the core and AXI-compliant data memory. The system bus decoder assigns distinct address spaces to each slave device (NVDLA and DRAM) to ensure efficient memory-mapped communication.

\begin{figure}[b!]
\vspace{-5mm}
\centering
\includegraphics[width=0.7\linewidth]{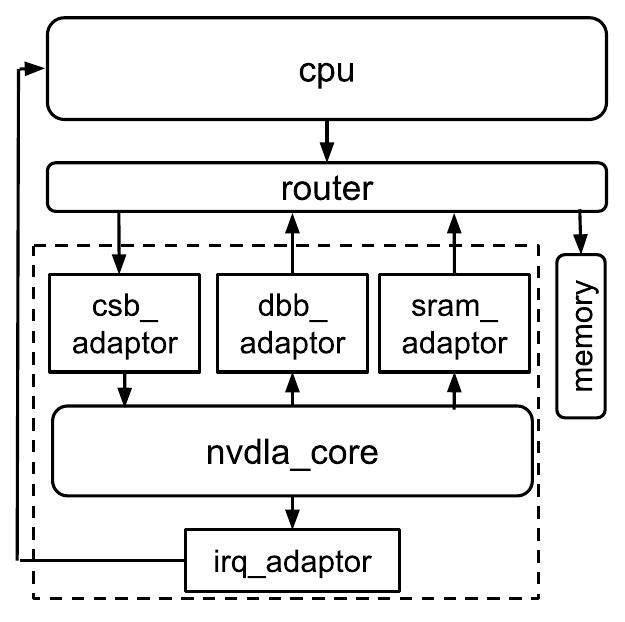}
\caption{NVDLA virtual platform.}
\label{fig:vp}
\end{figure}

\begin{figure*}[htbp]
  \centering
 \includegraphics[ width=0.95\textwidth]{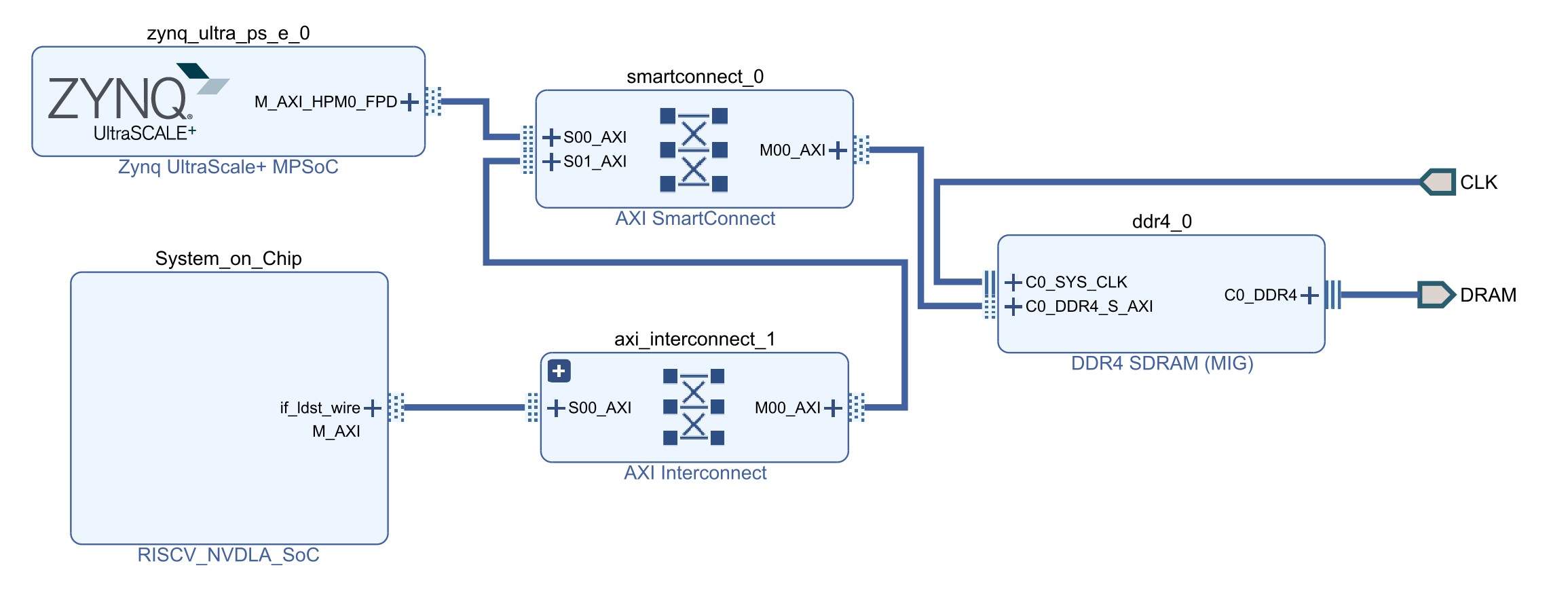}
  \caption{Set-up to test our SoC (Vivado block design of overall system set-up).}
  \label{fig:viv_bd}
\end{figure*}

\section{Methodology}
\label{sec:meth}

\subsection{Hardware Development Workflow}
\begin{enumerate}
    \item \textit{NVDLA Hardware Generation}: Parameterized Verilog code from the official NVDLA GitHub repository~\cite{git} is used to generate hardware configurations via the hardware tree build process, as outlined in the documentation~\cite{b2}.

\item \textit{System Integration}:  
    The NVDLA was integrated with a RISC-V processor using Codasip Studio. 
    A custom wrapper component was developed to encapsulate the NVDLA core, interface bridges, 
    and data-width converters, ensuring seamless compatibility. 
    The system bus, arbiter, and memory were interconnected through Codasip Studio's 
    \verb|testbench| construct to generate synthesizable RTL, which was subsequently imported into Vivado design suite
    (Fig.~\ref{fig:soc}). 
    The system bus decoder allocates two dedicated address spaces for the slave devices:
    \begin{itemize}
        \item \textit{NVDLA:} Address range \texttt{0x0 -- 0xFFFFF}, covering all configuration register addresses of the NVDLA
        \item \textit{DRAM:} Address range \texttt{0x100000 -- 0x200FFFFF}, providing access to 512~MB of DRAM data memory
    \end{itemize}

   This memory mapping enables the RISC-V processor to program the NVDLA using its standard load 
and store instructions for writing configuration registers and reading their status, 
eliminating the need for custom RISC-V instructions. 
The arbiter component coordinates DRAM access between the NVDLA (via its DBB interface) and 
the RISC-V processor (via its AHB interface), ensuring mutual exclusion and efficient 
memory utilization. 
This tightly coupled hardware interface enables bare-metal assembly programming for neural network execution.

\item \textit{Simulation and FPGA Prototyping}:  
    Behavioral simulation was performed in Vivado using RTL from the previous step along with Vivado IP cores for bridges and converters, while software binaries and neural network weights were loaded into memory. 
    After successful simulation, the design was synthesized and deployed on the FPGA board, utilizing the onboard DDR memory for input and weight storage. 
    Various DNN models were executed to evaluate system performance.

\end{enumerate}

\subsection{Software Development Flow}
\label{subsec:sw}

\setcounter{footnote}{0}
The software flow generates RISC-V machine code and extracts neural network weights from a Caffe model (Fig.~\ref{fig:system}). The Github repository\footnote{https://github.com/vineetbitsp/riscv-nvdla-sw} provides Python scripts, Linux commands, and detailed instructions for generating bare-metal RISC-V software through the following steps:

\begin{enumerate} 
\item \textit{Execution on Virtual Platform}: The Caffe model is compiled using the NVDLA compiler and executed on NVDLA’s VP, which provides a cycle-accurate co-simulation environment using QEMU and SystemC \cite{b2}. Interface-level transactions (CSB, DBB) are logged during execution (Fig~\ref{fig:vp}).

\item \textit{Configuration File Generation}:  
A Python script processes the VP log file by extracting lines containing the keyword \texttt{nvdla.csb\_adaptor}. Each entry represents a register transaction, categorized as read or write based on the \texttt{iswrite} flag:

\begin{itemize}
    \item Read operations (\texttt{iswrite=0}) are converted into \texttt{read\_reg} commands, which store the expected register values.
    \item Write operations (\texttt{iswrite=1}) are converted into \texttt{write\_reg} commands, specifying the target register address and the corresponding data value.
\end{itemize}
The generated command sequence constitutes the configuration file, which is subsequently converted into RISC-V assembly code. The assembly code is compiled into machine code using the RISC-V core SDK in \textit{Codasip Studio} and loaded into program memory for execution.

\item \textit{Weight Extraction}: To extract neural network weights, the python script filters VP log entries containing the keyword \texttt{nvdla.dbb\_adaptor}. Each entry corresponds to a data transaction:

\begin{itemize}
    \item Read operations (\texttt{iswrite=0}) indicate memory fetches, which correspond to weights.
    \item Write operations (\texttt{iswrite=1}) specify addresses and values being written to memory.
\end{itemize}
 Finally, duplicate address entries in the weight file are deleted by retaining the first occurrence, as they are the original weights. 
\end{enumerate}

\begin{table*}[htbp]
\centering
\caption{FPGA resource utilization (AMD's ZCU102 evaluation board) }
%\begin{adjustbox}{width=\textwidth}
\begin{tabular}{l|
>{\centering\arraybackslash}p{1.6cm}
>{\centering\arraybackslash}p{1.6cm}
>{\centering\arraybackslash}p{1.2cm}
>{\centering\arraybackslash}p{1.2cm}
>{\centering\arraybackslash}p{1.2cm}
>{\centering\arraybackslash}p{1.2cm}
>{\centering\arraybackslash}p{1.5cm}
>{\centering\arraybackslash}p{1.2cm}}
\hline
\multirow{2}{*}{\textbf{Major Components}} & \textbf{CLB LUTs} & \textbf{CLB Regs} & \textbf{CARRY8} & \textbf{F7 Muxes} & \textbf{F8 Muxes} & \textbf{CLBs} & \textbf{BRAM Tiles} & \textbf{DSPs} \\
(FPGA) & (274080) & (548160) & (34260) & (137040) & (68520) & (34260) & (912) & (2520) \\
\hline
Overall System Set-up (Fig.~\ref{fig:viv_bd}) & 96733 & 102823 & 1825 & 3719 & 1133 & 19898 & 323.5 & 39 \\
\quad MIG DDR4                  & 8651  & 10260  & 56   & 164  & 0    & 1754  & 25.5  & 3  \\
\quad AXI SmartConnect          & 5546  & 7860   & 0    & 0    & 0    & 1137  & 0     & 0  \\
\quad Our SoC (Fig.~\ref{fig:soc})                  & 81986 & 83659  & 1762 & 3555 & 1133 & 17025 & 298   & 36 \\
\quad\quad nv\_small NVDLA           & 74575 & 79567  & 1569 & 3091 & 1048 & 15734 & 66    & 32 \\
\quad\quad uRISC\_V core             & 6346  & 2767   & 173  & 419  & 67   & 1297  & 0     & 4  \\
\quad\quad Program Memory            & 241   & 6      & 0    & 45   & 18   & 148   & 232   & 0  \\
\hline
\end{tabular}
%\end{adjustbox}
\label{tab:fpga_utilization}
\end{table*}

\begin{table}[h!]
    \centering
    \caption{Evaluation of our SoC containing nv\_small NVDLA (FPGA implementation results)}
    \label{tab:perf_eval}
    \resizebox{\columnwidth}{!}{ % Resize to fit within one column
    \begin{tabular}{l|c c c c c}
        \hline
        \textbf{Model} & \textbf{Layers} & \textbf{Input} & \makecell{\textbf{Model}\\\textbf{Size}} 
        & \makecell{\textbf{Proc. Time}\\\textbf{@100MHz}} 
        & \makecell{\textbf{Proc. Time}\\\textbf{@50MHz}~\cite{b6}} \\ 
        \hline
        LeNet-5 & 9  & 1$\times$28$\times$28 & 1.7 MB & \textbf{4.8 ms} & \text{263 ms} \\ 
        
        ResNet-18 & 86 & 3$\times$32$\times$32 & 0.8 MB & \textbf{16.2 ms} & \text{NA} \\ 
        
        ResNet-50 & 228 & 3$\times$224$\times$224 & 102.5 MB & \textbf{1.1 s} & \text{2.5 s} \\ 
        \hline
    \end{tabular}
    }
\end{table}

\begin{table}[h!]
\centering
\caption{Evaluation of our SoC containing nv\_full NVDLA (Simulation results)}
\resizebox{\columnwidth}{!}{%
\begin{tabular}{l| c c c c}
\hline
\textbf{Model} & \shortstack{\textbf{Input} \\ \textbf{size}} & \shortstack{\textbf{Model} \\ \textbf{size}} & \shortstack{\textbf{Number of} \\ \textbf{clock cycles}} & \shortstack{\textbf{Processing time} \\ \textbf{@100 MHz (ms)}} \\
\hline
LeNet-5     & 1x28x28    & 1.7 MB    & 143188     & 1.4   \\
% \hline
ResNet-18   & 3x32x32    & 813.5 KB  & 324387     & 3.2   \\
% \hline
ResNet-50   & 3x224x224  & 102.5 MB  & 26565315   & 265   \\
% \hline
MobileNet   & 3x224x224  & 17 MB     & 22525704   & 220   \\
% \hline
GoogleNet   & 3x224x224  & 53.5 MB   & 40889646   & 408   \\
% \hline
AlexNet     & 3x227x227  & 243.9 MB  & 35535582   & 355   \\
\hline
\end{tabular}
}
\label{tab:nv_full}
\end{table}

\begin{comment}
\begin{table}[h]
    \centering
    \caption{Defines for NVDLA RTL to be specified in Vivado project}
    \label{tab:defines}
    \renewcommand{\arraystretch}{1} % Increase row height for readability
    
     \fontsize{16pt}{22pt}\selectfont % Adjust font size (10pt with 12pt line spacing)
    \resizebox{\columnwidth}{!}{ % Adjust table to fit within column width
        \begin{tabular}{l | l }
            \hline
            \\
            \makecell{\textbf{NVDLA} \\ \textbf{Configuration}} & \textbf{List of Defines} \\
            \\
            \hline
            \\
            nv\_small & \makecell[l] {SYNTHESIS \\ DESIGNWARE\_NOEXIST \\ VLIB\_BYPASS\_POWER\_CG \\ NV\_FPGA\_FIFOGEN \\ FPGA \\ FIFOGEN\_MASTER\_CLK\_GATING\_DISABLED} \\
            \\
            
            nv\_full & \makecell[l]{SYNTHESIS \\ DESIGNWARE\_NOEXIST} \\
            \\
            \hline
        \end{tabular}
    }
\end{table}

\end{comment}

\section{Evaluation and Testing}
\label{sec:perf}

This section presents the performance evaluation of the proposed SoC, including its FPGA implementation and testing with standard neural network models. Initial functional validation was performed via behavioral simulation using standard NVDLA test traces such as sanity, convolution and memory tests available from the NVDLA Github repository. These were translated into RISC-V assembly and used to verify the correctness of the integrated SoC design. 

To support larger models, external DRAM was connected to the SoC through a DDR4 memory controller (MIG DDR4), and is initialized via the ARM core of the Zynq UltraScale+ MPSoC on the ZCU102 board. This configuration enables access to 512 MB of DDR4 memory from the programmable logic. The system architecture was implemented in AMD Vivado, with a high-level interface diagram shown in Fig.~\ref{fig:viv_bd}. The Zynq core initializes the DRAM with both the weight file and input image. At any given time, the DRAM is connected either to the Zynq core or the SoC using an AXI SmartConnect, which functions as a multiplexer. Additionally, an AXI Interconnect is placed between the SoC and MIG DDR4 to reconcile frequency mismatches, since the SoC operates at 300 MHz while the DDR4 runs at 100 MHz. The complete block design was synthesized and deployed on the FPGA. 

The SoC was successfully tested with standard deep learning models, including LeNet-5, ResNet-18, and ResNet-50. During execution, the DRAM is preloaded with weight and image files in \texttt{.bin} format. The RISC-V program memory, implemented using FPGA block RAMs, is loaded with machine code generated from the configuration file in \texttt{.mem} format.

While Table~\ref{tab:fpga_utilization} shows the FPGA resource utilization for the complete system set-up, our SoC, and its major components, Table~\ref{tab:perf_eval} reports the execution times at a system clock frequency of 100\,MHz. The execution speed outperforms previous work, where NVDLA was integrated on a 64-bit RISC-V--based platform, as shown in the table. Table~\ref{tab:nv_full} presents simulation results for the \texttt{nv\_full} configuration of NVDLA, including total cycle counts and processing times at 100\,MHz. Although the \texttt{nv\_full} configuration delivers higher performance than \texttt{nv\_small}, it is an enormous design and does not fit on most FPGAs, including the ZCU102 FPGA board used in this work. For this device, the LUTs overutilization was quite substantial for \texttt{nv\_full} as observed during synthesis. 

The \texttt{nv\_small} configuration supports only INT8 precision, while \texttt{nv\_full} additionally supports FP16 computations. The models included in the Table~\ref{tab:nv_full} shows computation times with FP16 precision. The performance comparison of ResNet-50 on both configurations highlights that \texttt{nv\_full} is substantially faster, as it integrates a larger number of MAC units. A limitation of this work is that the \texttt{nv\_small} configuration currently supports only a limited set of models, primarily due to the lack of INT8 calibration tables. Future work will address this limitation to broaden model support.

\section*{Future Work}
Future development will focus on extending model support for the nv\_small configuration to include additional deep learning models such as MobileNet, GoogleNet, and AlexNet. Two promising directions are:
\begin{enumerate}
    \item Generating INT8 calibration tables required by the NVDLA compiler, which are not currently provided but are partially described in the NVDLA GitHub documentation~\cite{git}.

    \item Integrating the ONNC compiler~\cite{git-onnc} to generate NVDLA-compatible loadable files from ONNX models, enabling broader deployment through execution on the NVDLA VP.
\end{enumerate}

\section{Conclusion}

This work presents a custom SoC integrating the NVDLA accelerator with a RISC-V processor, implemented and validated on an FPGA platform. The current design leverages the nv\_small configuration, with the flexibility to support nv\_full by modifying parameters such as the AXI interface width (e.g., from 64-bit to 512-bit). The SoC operates without the need for a Linux kernel, enabling a lightweight, standalone execution model ideal for edge AI applications requiring low latency and constrained resources. FPGA synthesis results demonstrate the feasibility of this design on low- to mid-range devices.

\vspace{12pt}
\color{red}

\end{document}